\documentclass[showpacs, preprintnumbers, amsmath, amssymb, floatfix, twocolumn]{revtex4}
\usepackage{graphicx}
\usepackage{dcolumn}
\usepackage{bm}
\usepackage{color}%

\begin{document}

\title{Robust tunability of magnetorestance in Half-Heusler $R$PtBi ($R$ = Gd, Dy, Tm, and Lu) compounds}
\author{Eundeok Mun$^{1,2}$, Sergey L. Bud'ko$^{1}$, Paul C. Canfield$^{1}$}
\affiliation{$^{1}$ Ames Laboratory US DOE and Department of Physics and Astronomy, Iowa State University, Ames, IA 50011, USA}
\affiliation{$^{2}$ Department of Physics, Simon Fraser University, Burnaby, BC V5A1S6, Canada}


\begin{abstract}
We present the magnetic field dependencies of transport properties for $R$PtBi ($R$ = Gd, Dy, Tm, and Lu) half-Heusler compounds. Temperature and field dependent resistivity measurements of high quality $R$PtBi single crystals reveal an unusually large, non-saturating magnetoresistance (MR) up to 300 K under a moderate magnetic field of $H$ = 140 kOe. At 300 K, the large MR effect decreases as the rare-earth is traversed from Gd to Lu and the magnetic field dependence of MR shows a deviation from the conventional $H^{2}$ behavior. The Hall coefficient ($R_{H}$) for $R$ = Gd indicates a sign change around 120 K, whereas $R_{H}$ curves for $R$ = Dy, Tm, and Lu remain positive for all measured temperatures. At 300 K, the Hall resistivity reveals a deviation from the linear field dependence for all compounds. Thermoelectric power measurements on this family show strong temperature and magnetic field dependencies which are consistent with resistivity measurements. A highly enhanced thermoelectric power under applied magnetic field is observed as high as $\sim$100 $\mu$V/K at 140 kOe. Analysis of the transport data in this series reveals that the rare-earth-based Half-Husler compounds provide opportunities to tune MR effect through lanthanide contraction and to elucidate the mechanism of non-trivial MR.
\end{abstract}



\maketitle

\section{Introduction}

The faced-centered cubic $R$PtBi ($R$ = rare-earth) family is a stoichiometric Half-Heusler system. Initial resistivity measurements showed that the resistivity systematically changes from a small gap semiconductor for lighter rare-earth compounds to metallic (or semimetallic) for heavier rare-earth compounds \cite{Canfield1991}. Specific members of the series have been previously characterized in some detail: YbPtBi for its heavy fermion behavior and field tuned quantum criticality in YbPtBi \cite{Canfield1991, Fisk1991, Mun2013}, CePtBi for magnetic field induced Lifshitz transition  \cite{Goll2002, Wosnitza2006}, and the non moment bearing $R$PtBi (R = Y, La, and Lu) for non-trivial superconductivity \cite{Butch2011, Goll2008, Tafti2013}. In addition, the combination of strong spin-orbit coupling and non-centrosymmetric crystal structure (absence of an inversion center) makes Half-Heusler compounds a strong candidate for 3-dimensional topological materials \cite{Chadov2010, Lin2010, Xiao2010}. Considerable efforts have been expended on experimental studies of the electronic structures and transport properties of Y- and Lu-based Half-Heusler materials \cite{Gofryk2011, Shekhar2012a, Shekhar2012b, Wang2012}. In addition, the specific antiferromagnetic (AFM) ordering in GdPtBi may lead to an even more exotic state, proposed to be the AFM topological insulator \cite{Kreyssig2011, Muller2014}. Although the confirmation of the topological properties in this family remains unresolved, the validity of the band structure calculations has been brought into question by recent angle resolved photoemission spectroscopy (ARPES) experiments on $R$PtBi compounds, which found surface dispersions that differ from those calculated in bulk and showed a clear spin-orbit splitting of the surface bands that cross the Fermi surface \cite{Liu2011}.

Recently, large, linear field dependence of magnetoresistance (MR) without any sign of saturation has been observed in a number of materials such as Bi thin films, narrow-band gap semiconductors (silver chalcogenides Ag$_{2+\delta}$Te and Ag$_{2+\delta}$Se), doped-InSb, and Dirac electron systems (graphene and topological insulators) \cite{Yang1999, Xu1997, Hu2008, Friedman2010, Thio1998, Butch2010}. Although, the observed anomalous MR in many of these materials has been explained with linearly dispersing excitation, mobility fluctuation, and surface state, the mechanism of this anomalous MR is not fully understood. In addition, exceptionally large MR was found in semimetal PtSn$_{4}$ \cite{Mun2012} and then later rediscovered in WTe$_{2}$ \cite{ Ali2014} and NbSb$_{2}$ \cite{Wang2014}.

In this paper we present the results of magnetization and transport measurements on single crystalline samples of $R$PtBi ($R$ = Gd, Dy, Tm, and Lu). We report anomalous and large magnetotransport properties, especially showing a large MR effect even at room temperatures. Transport measurements of high quality $R$PtBi single crystals reveal that a large, non-saturating MR as high as $\sim$800 \% at 150 K and $\sim$300 \% at 300 K at $H$ = 140 kOe. The magnetic field dependence of MR deviates significantly from the conventional $H^{2}$ behavior. The size of MR effect in this family is comparable or smaller than that of high-quality, ordinary metals at low temperatures, but the MR is much larger than the same high purity metals at high temperatures. These extraordinary high temperature MR effects can be systematically controlled by varying rare-earth elements without introducing disorder.

\begin{figure}
\centering
\includegraphics[width=1\linewidth]{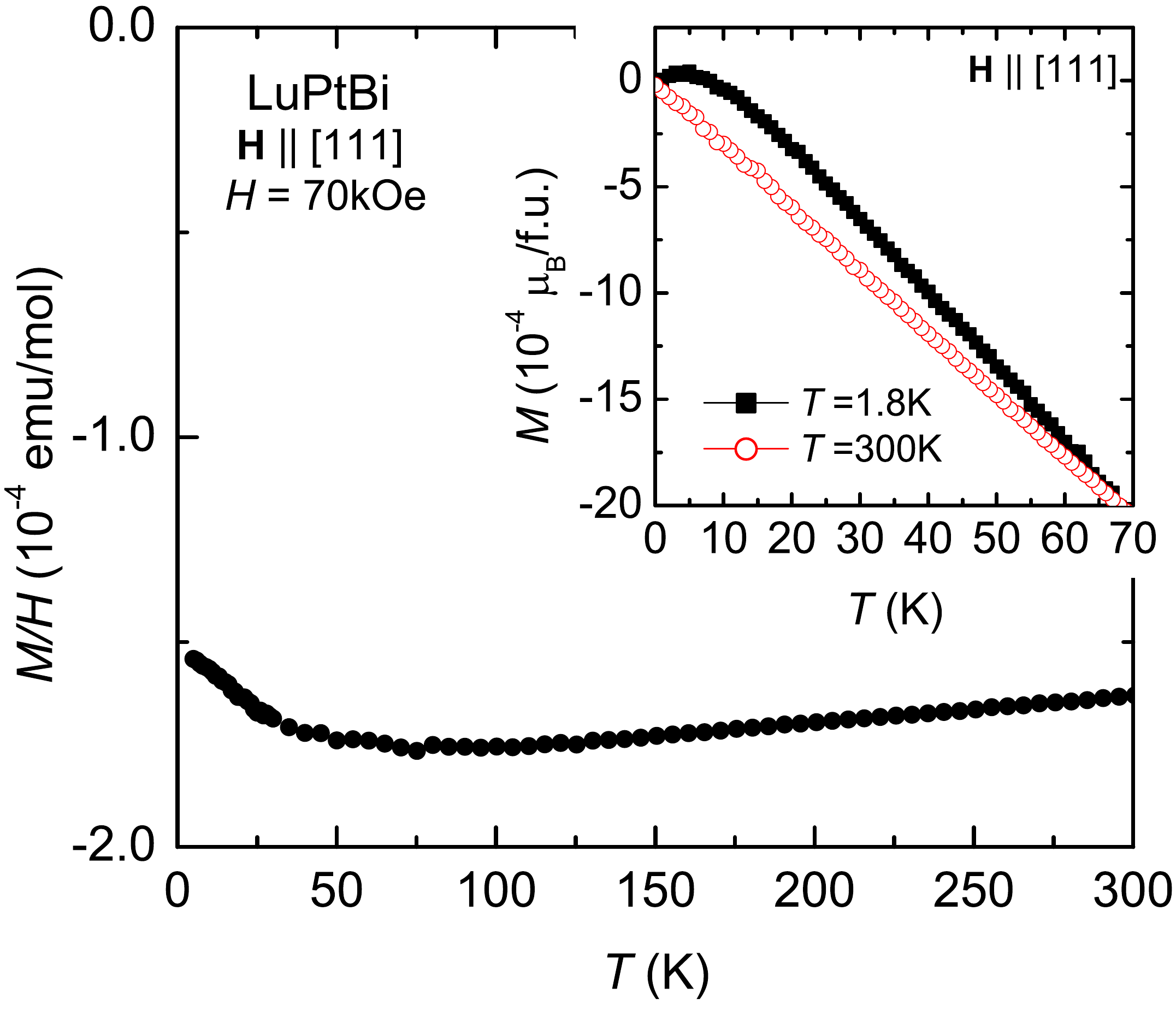}
\caption{$M(T)/H$ data for LuPtBi for $H$ = 70 kOe. Inset: Magnetization isotherms of LuPtBi at $T$ = 1.8 K and 300 K.}
\label{Fig1}%
\end{figure}

\section{Experimental}

Single crystals of $R$PtBi ($R$ = Gd, Dy, Tm, and Lu) were grown out of ternary melt with excess of Bi \cite{Canfield1991, Canfield1992}. The lattice parameters of these cubic structures, with space group F$\bar{4}$3m, were determined by powder x-ray diffraction measurements, and are consistent with previous values \cite{Canfield1991}. The dc magnetization as a function of temperature from 1.8 to 300 K, and magnetic field, up to 70 kOe, was measured in a Quantum Design (QD) Magnetic Property Measurement System (MPMS). Four-probe ac resistivity measurements as a function of temperature, $\rho(T)$, and magnetic field, $\rho(H)$, were performed from 300 K down to 2 K in a QD Physical Property Measurement System (PPMS). It is a well known fact that thin film Bi produces a large MR effect even up to room temperature \cite{Yang1999}. In order to avoid any contribution from residual Bi flux, forming thin films on the sample surface, all surfaces of resistivity bars parallel to the current flow were polished. The Hall resistivity, $\rho_{H}$, was measured in a standard 4 probe configuration, by reversing the magnetic field to eliminate the effects of small misalignment of the voltage wires. The thermoelectric power (TEP) as a function of temperature, $S(T)$, and field, $S(H)$, was measured using a dc, alternating heating, technique utilizing two heaters, and two thermometers \cite{Mun2010}. The magnetic field was applied along the [111] direction for all measurements, and the electric current (\textbf{I}) and the temperature gradient ($\Delta T$) across the sample was applied perpendicular to [111]: \textbf{H} $\parallel$ [111], \textbf{I} $\perp$ [111], and $\Delta T$ $\perp$ [111].

\begin{figure}
\centering
\includegraphics[width=1\linewidth]{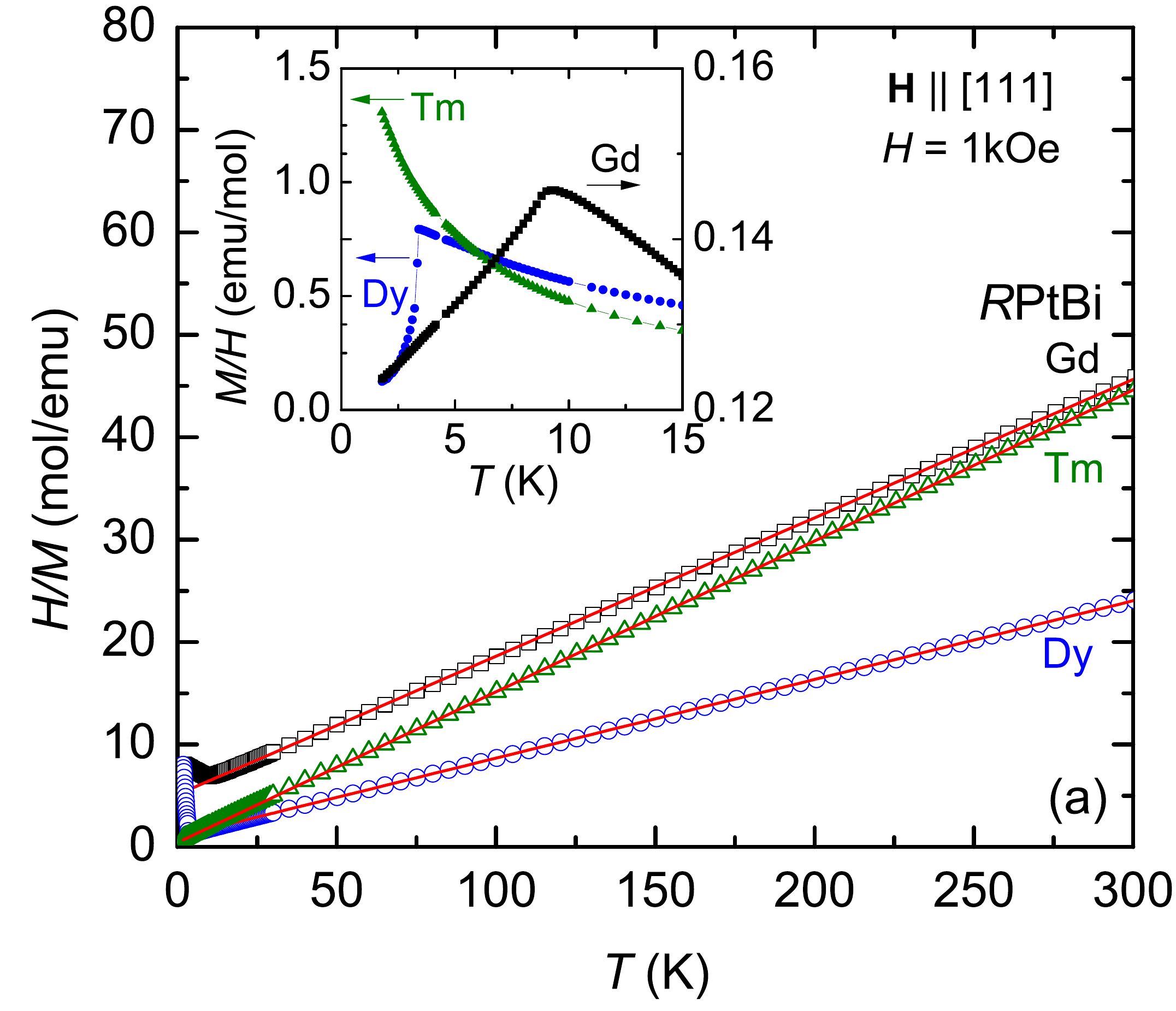}
\includegraphics[width=1\linewidth]{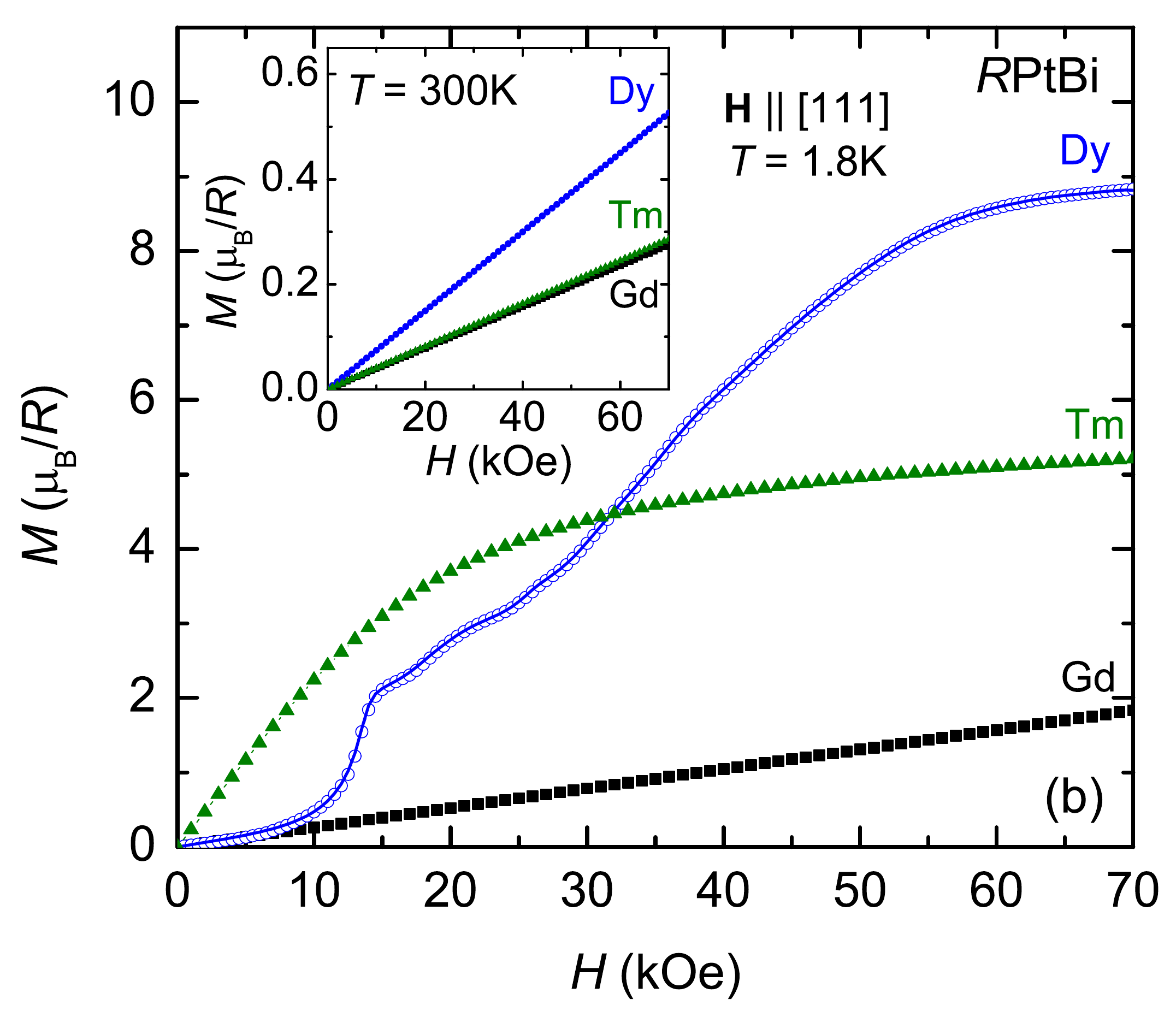}
\caption{(a) $H/M(T)$ data for $R$PtBi ($R$ = Gd, Dy, and Tm), taken at $H$ = 1 kOe. Solid lines represent Curie-Weiss fits. Inset shows $M(T)/H$ curves below 15 K. (b) Magnetization isotherms of $R$PtBi ($R$ = Gd, Dy, and Tm) at $T$ = 1.8 K and 300 K (inset)}
\label{Fig2}%
\end{figure}

\begin{figure}
\centering
\includegraphics[width=1\linewidth]{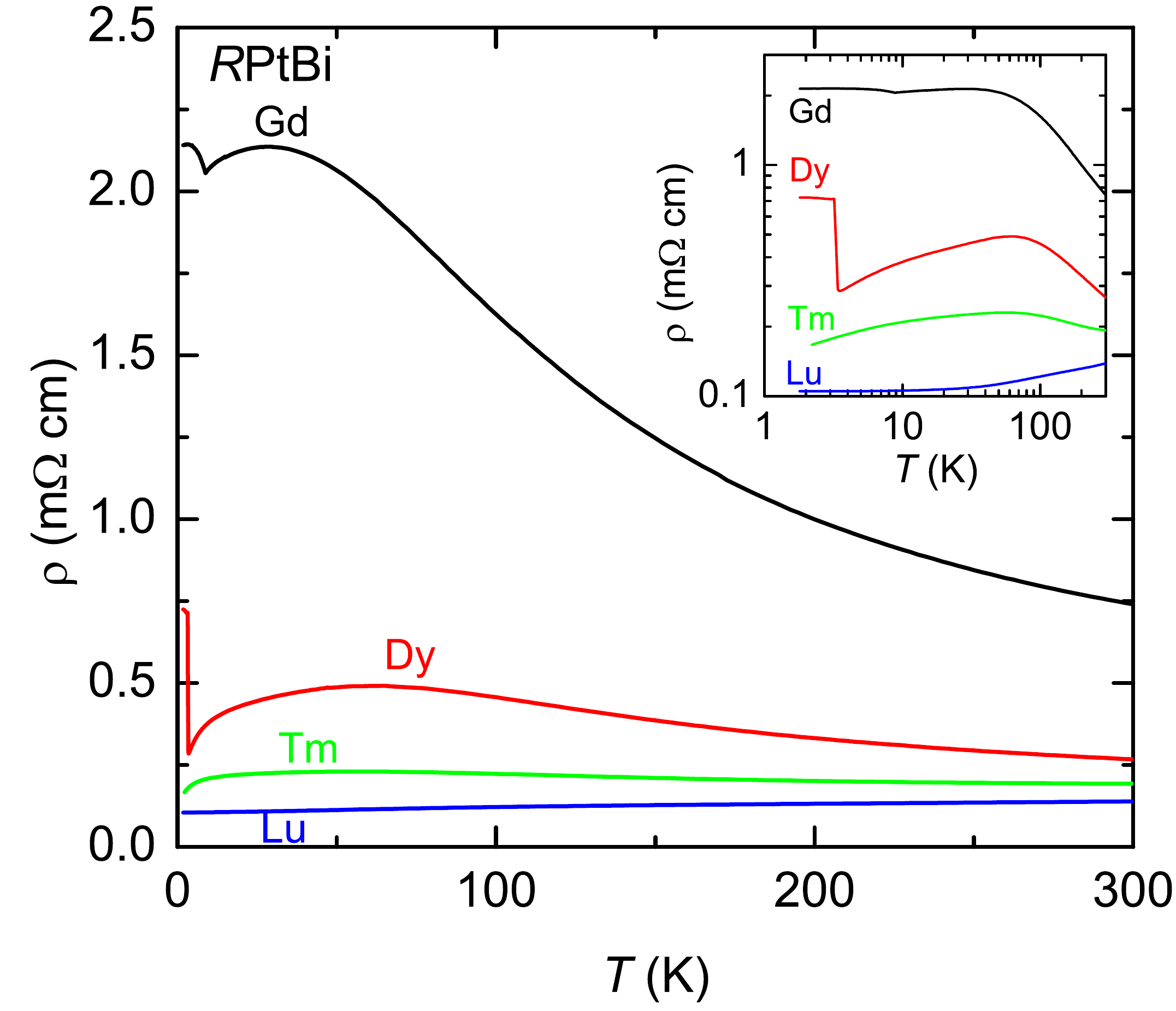}
\caption{Electrical resistivity $\rho(T)$ of $R$PtBi ($R$ = Gd, Dy, Tm, and Lu) at $H$ = 0. Inset presents a log-log plot of the $\rho(T)$ data.}
\label{Fig3}%
\end{figure}

\section{Results}

The $M(T)/H$ data of LuPtBi for $H$ = 70 kOe are essentially temperature independent as shown in Fig. \ref{Fig1}. The magnetization isotherm, $M(H)$, for $T$ = 300 K shows a linear magnetic field dependence up to 70 kOe, as shown to the inset. A small upturn in the magnetic susceptibility at low temperatures and a small hump in the $T$ = 1.8 K magnetization isotherm (see inset) at low fields are probably associated with small amounts of paramagnetic impurities (most likely trace amounts of heavy rare-earths on the Lu-site). The $H / M(T)$ data measured at $H$ = 1 kOe, for $R$PtBi ($R$ = Gd, Dy, and Tm) single crystals are presented in Fig. \ref{Fig2} (a). The solid lines are fits to the data assuming Curie-Weiss behavior and yield an effective moments close to those of the trivalent states of the rare-earth ions. The observed antiferromagnetic (AFM) phase transitions, shown in the inset, are consistent with the earlier report in Ref. \cite{Canfield1991}. The AFM transition temperatures determined from $d (M/H) \cdot T / dT$ are: $T_{N}$ = 8.8 K for Gd and $T_{N}$ = 3.3 K for Dy. The $M(H)$ curves for $R$PtBi ($R$ = Gd, Dy, and Tm) at $T$ = 1.8 K are plotted in Fig. \ref{Fig2} (b). For GdPtBi, $M(H)$ increases linearly and reaches $\sim$ 2 $\mu_{B}$/Gd$^{3+}$ at 70 kOe, which is far below the theoretical saturated value of 7 $\mu_{B}$ for the free Gd$^{3+}$ ion, suggesting a strong AFM interaction. The $M(H)$ curve for DyPtBi clearly shows step-like-jumps (metamagnetic transitions) below $T_{N}$. A broad enhancement of $M(H)$ for TmPtBi, following a Brillouin function-like behavior, is observed. Note that the magnetization values at 70 kOe for $R$ = Dy and Tm are lower than the theoretical values of 10 $\mu_{B}$ and 7 $\mu_{B}$ for the saturated moment of free Dy$^{3+}$ and Tm$^{3+}$ ions, respectively. At 300 K the $M(H)$ curves (inset) for all three compounds follow a linear behavior in magnetic fields up to 70 kOe.

At $H$ = 0, electric resistivity curves for $R$PtBi ($R$ = Gd, Dy, Tm, and Lu) are plotted in Fig. \ref{Fig3}. $\rho(T)$ for $R$ = Gd, Dy, and Tm increase as temperature decreases from 300 K, and then show a broad hump below 100 K. At the magnetic ordering temperature ($T_{N}$), the $\rho(T)$ curves for $R$ =  Gd and Dy show sharp rises as a signature of AFM ordering as shown in the inset, indicating the opening of a magnetic superzone gap. The $\rho(T)$ for LuPtBi reveals a metallic behavior below 300 K. The absolute value of resistivity of these $R$PtBi compounds decreases continuously as the rare-earth series is traversed from Gd to Lu.

\begin{figure}
\centering
\includegraphics[width=1\linewidth]{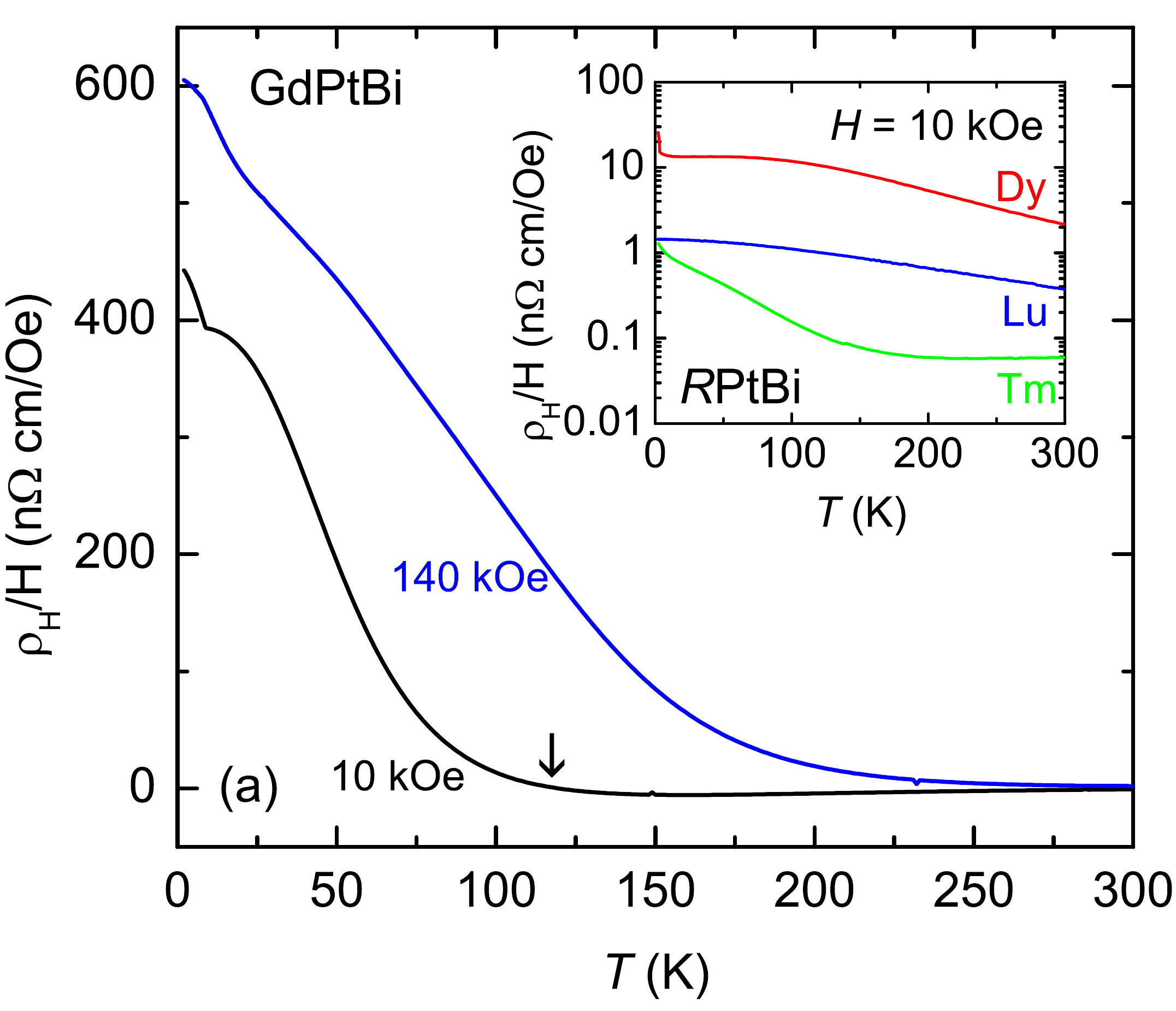}
\includegraphics[width=1\linewidth]{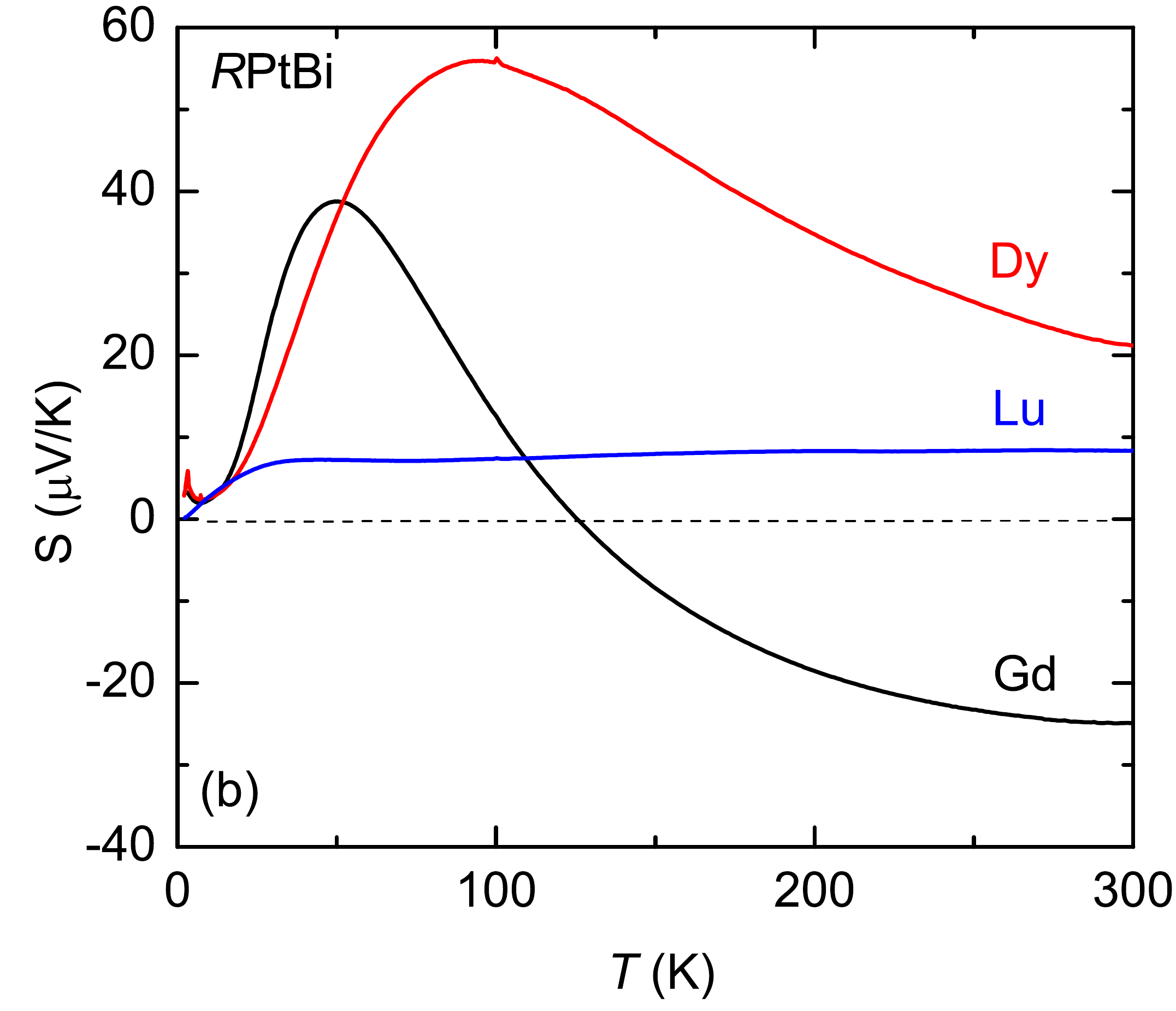}
\caption{(a) Hall coefficient $R_{H} = \rho_{H}/H$ of GdPtBi at $H$ = 10 and 140 kOe. The vertical arrow indicates a sign change. Inset shows $R_{H}$ for $R$ = Dy, Tm, and Lu at $H$ = 10 kOe. (b) Thermoelectric power $S(T)$ for $R$ = Gd, Dy, and Lu at $H$ = 0.}
\label{Fig4}%
\end{figure}

\begin{figure*}
\centering
\includegraphics[width=0.4\linewidth]{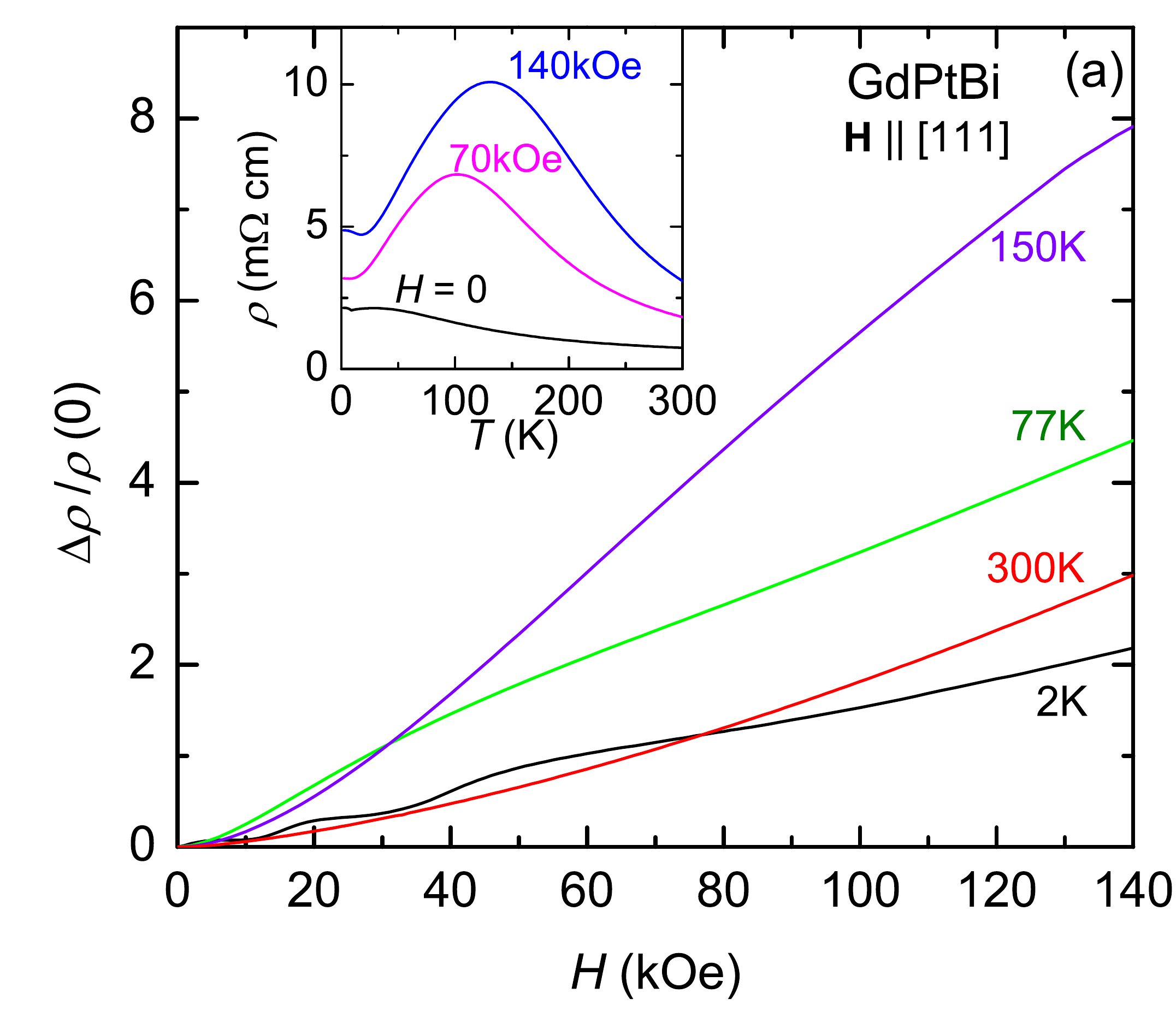}\includegraphics[width=0.4\linewidth]{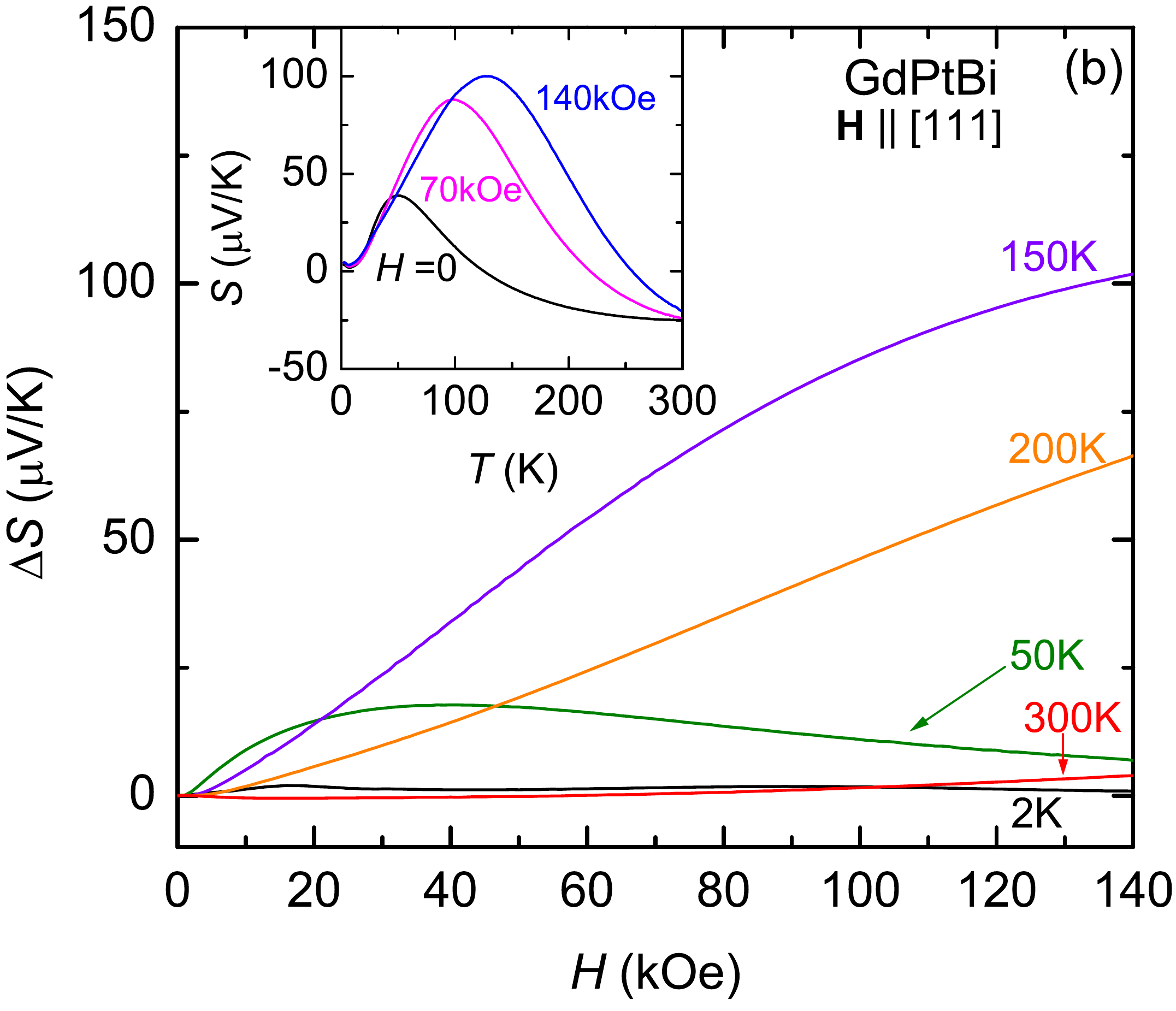}
\includegraphics[width=0.4\linewidth]{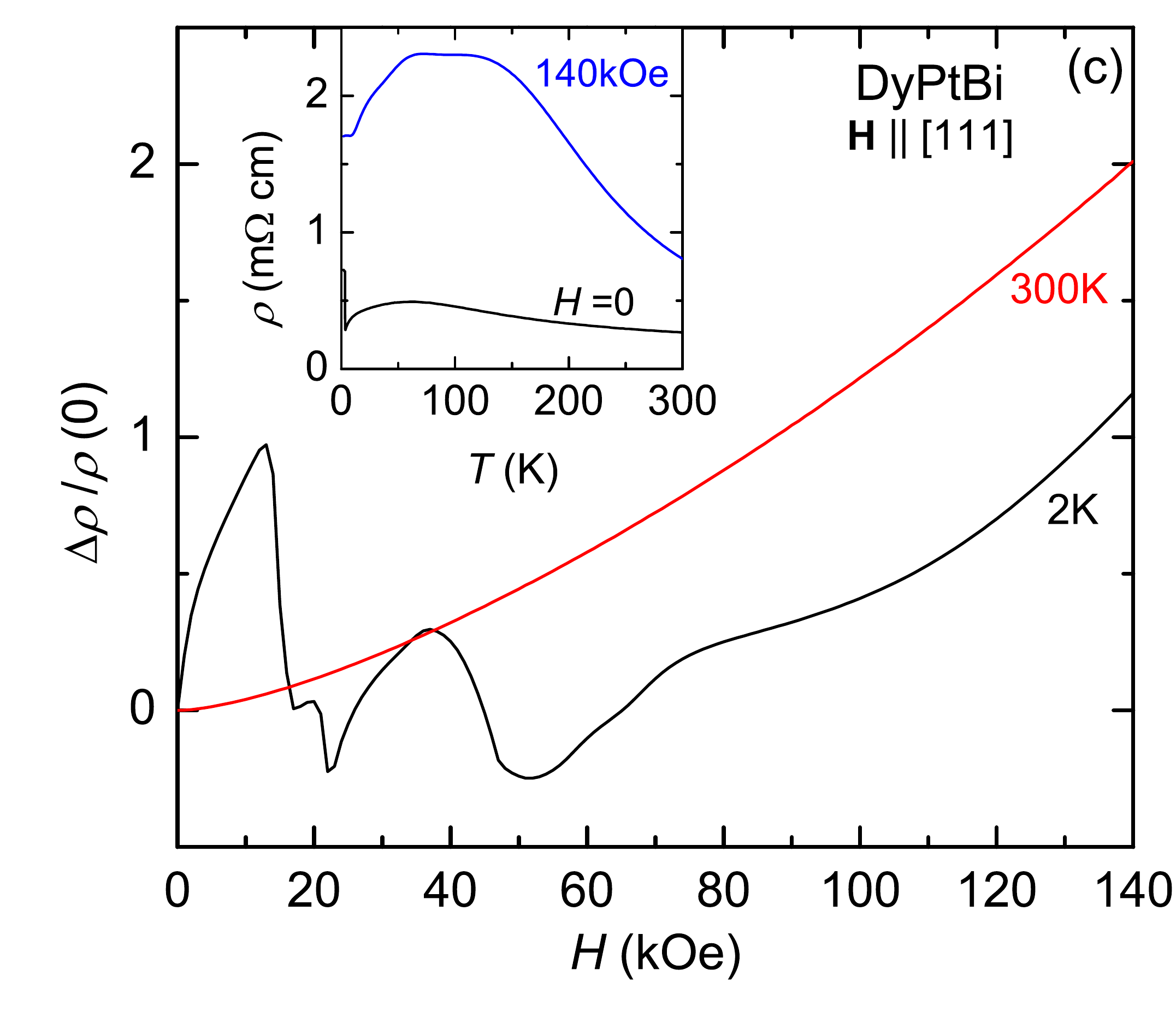}\includegraphics[width=0.4\linewidth]{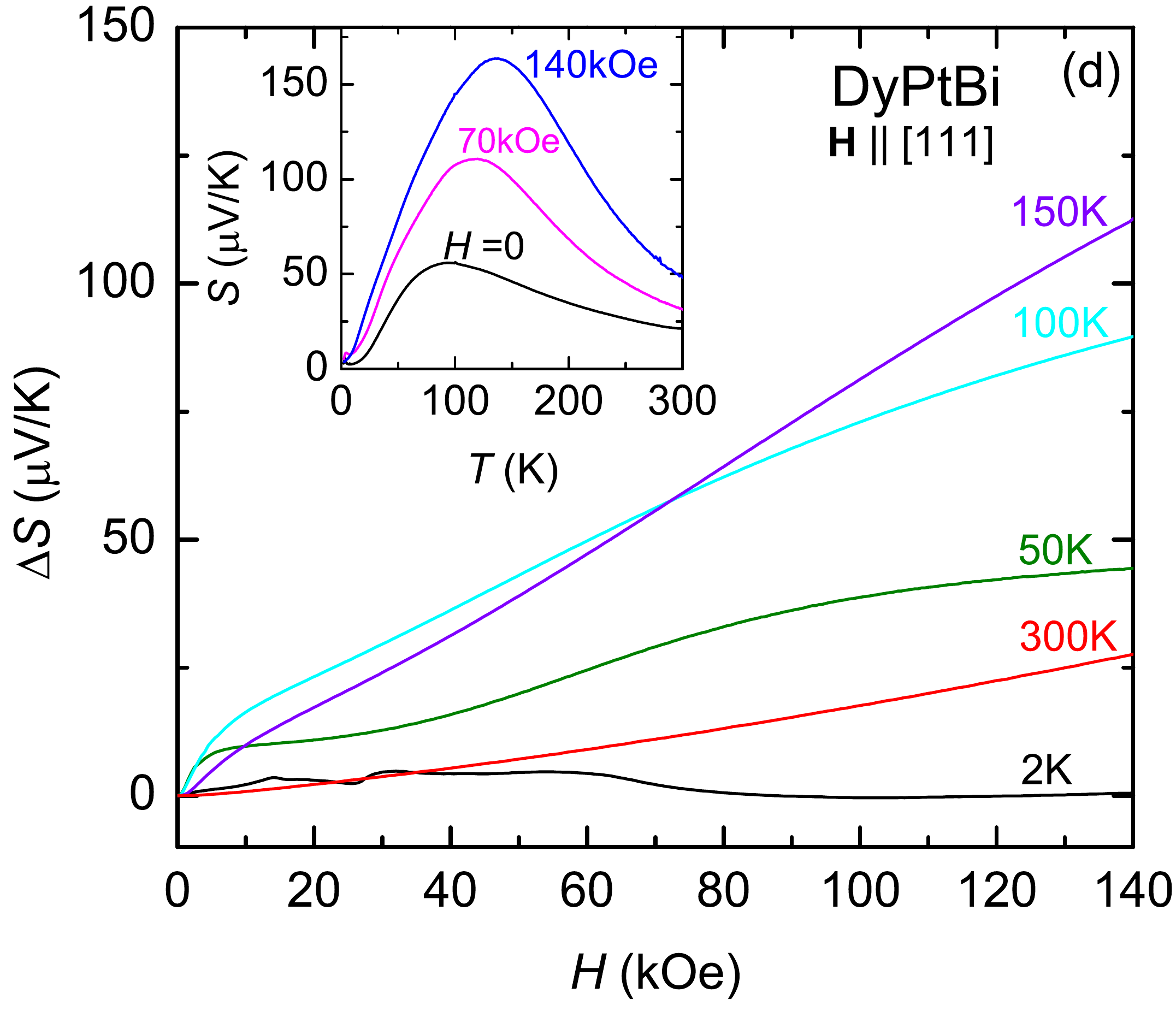}
\includegraphics[width=0.4\linewidth]{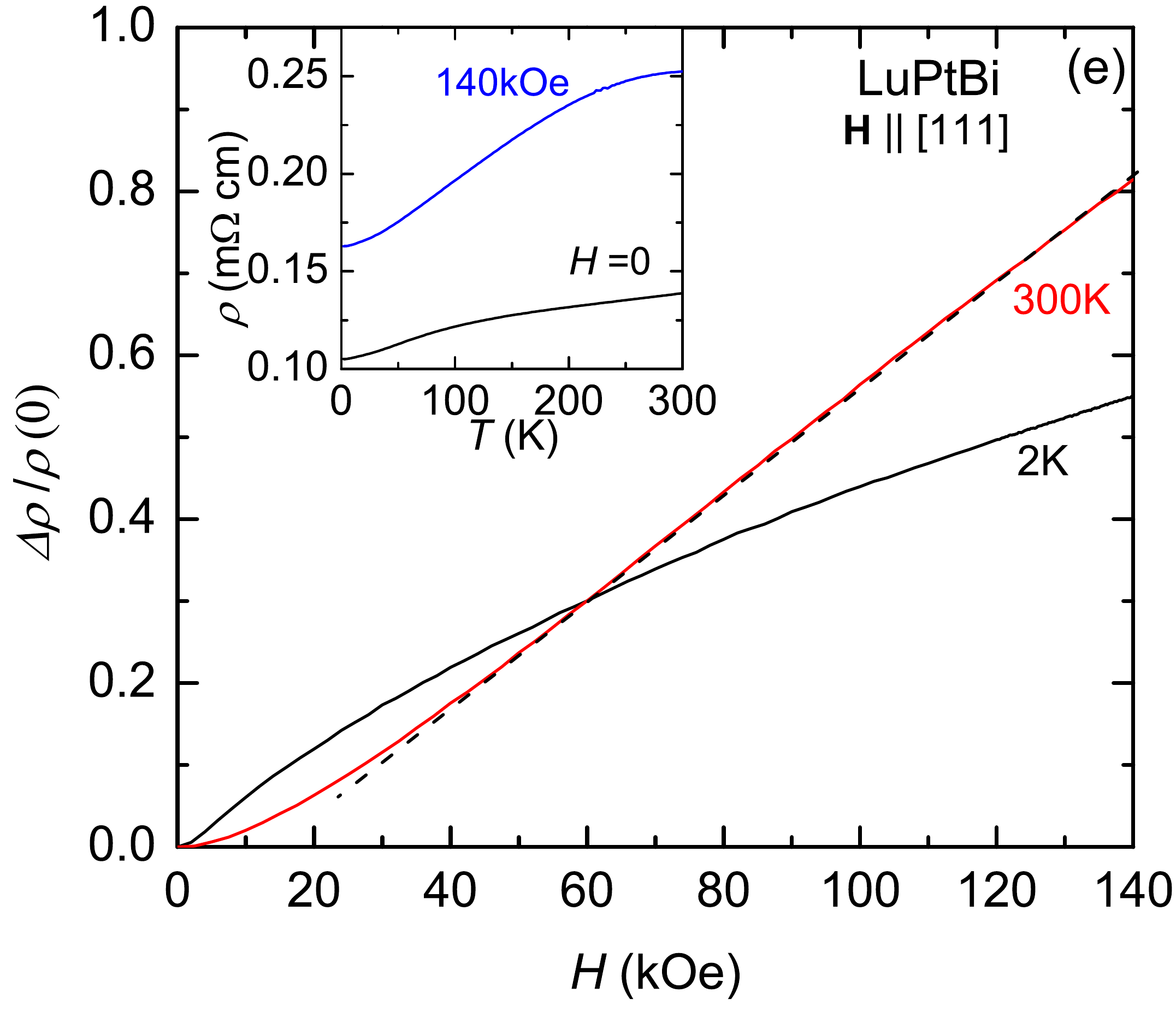}\includegraphics[width=0.4\linewidth]{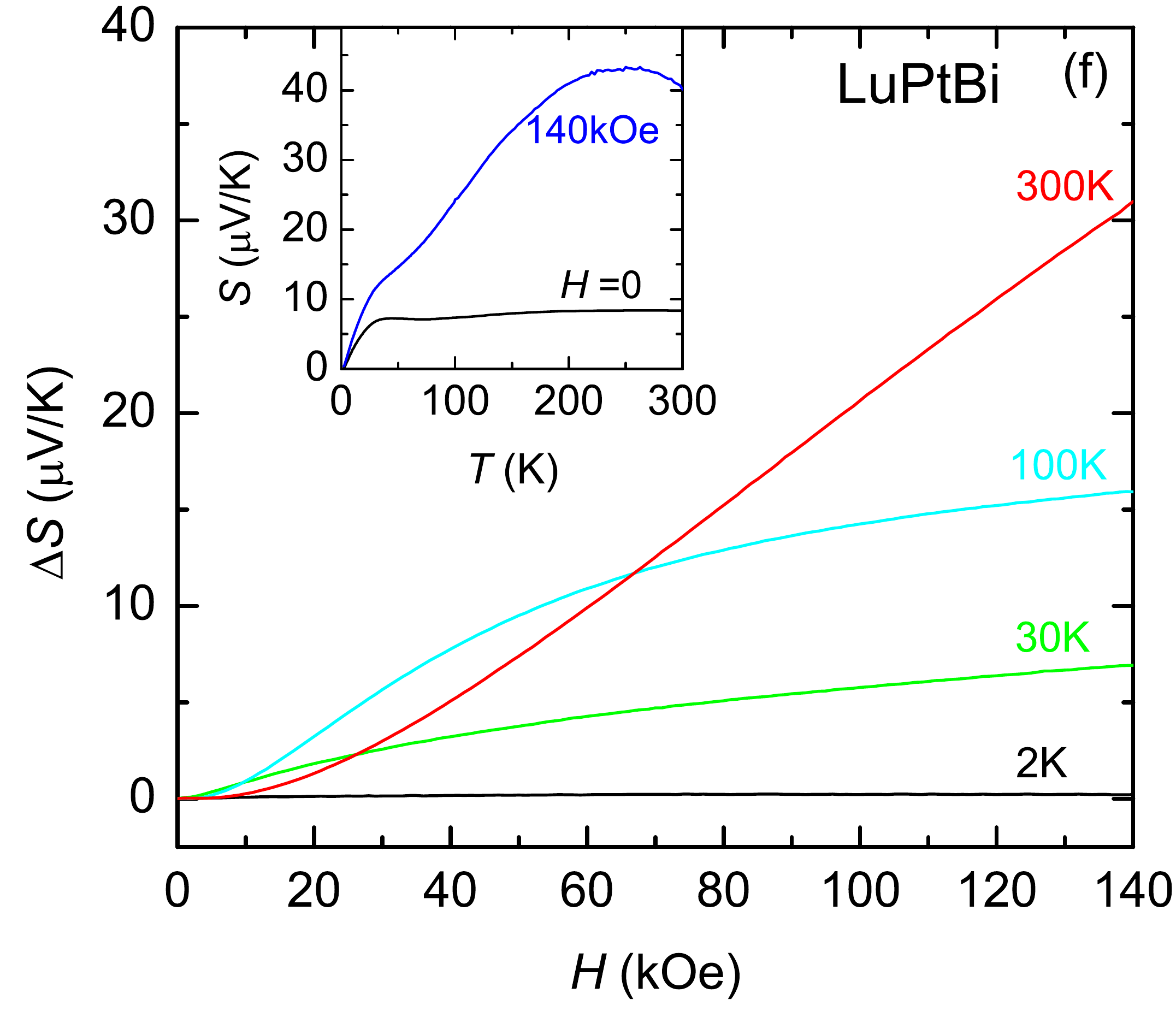}
\caption{Electrical resistivity $\rho(T, H)$ and thermoelectric power $S(T, H)$ for $R$PtBi ($R$ = Gd, Dy, and Lu). Magnetoresistance, $\Delta \rho/\rho(0) = [\rho(H) - \rho(0)]/\rho(0)$, of (a) $R$ = Gd, (c) $R$ = Dy, and (e) $R$ = Lu as a function of magnetic field at selected temperatures. Insets show electrical resistivity, $\rho(T)$, of (a) $R$ = Gd, (c) $R$ = Dy, and (e) $R$ = Lu as a function of temperature at selected magnetic fields. The dashed-line on top of the 300 K curve in (e) is guide to the eyes. Magneto thermoelectric power, $\Delta S = S(H) - S(0)$, of (b) $R$ = Gd, (d) $R$ = Dy, and (f) $R$ = Lu as a function of magnetic field at selected temperatures. Insets show thermoelectric power, $S(T)$, of (b) $R$ = Gd, (d) $R$ = Dy, and (f) $R$ = Lu as a function of temperature at selected magnetic fields.}
\label{Fig5}%
\end{figure*}

The temperature-dependent Hall coefficient, $R_{H} = \rho_{H}/H$, for GdPtBi is presented in Fig. \ref{Fig4} (a). The $R_{H}$ for $H$ = 10 kOe depends strongly on temperature accompanied with a sign reversal around 120 K; $R_{H}$ $<$ 0 for $T$ $>$ 120 K and $R_{H}$ $>$ 0 for $T$ $<$ 120 K. For $H$ = 140 kOe, $R_{H}$ rises more strongly below 200 K, and also remains positive up to 300 K. In zero applied field, a sign change in GdPtBi can be clearly seen in TEP measurement as shown in Fig. \ref{Fig4} (b). At high temperatures, $S(T)$ for $R$ = Gd is negative and increases with decreasing temperature, passing through the sign reversal from negative to positive at $\sim$127 K. At low temperatures, $S(T)$ has a broad peak around 50 K and clear kink at $T_{N}$. The $R_{H}$ curves for $R$ = Dy, Tm, and Lu are positive for all temperatures measured as shown in the inset of Fig. \ref{Fig4} (a), which are consistent with TEP data. The $S(T)$ for $R$ = Dy shows a similar behavior to $R$ = Gd, but retains a positive sign up to 300 K. The $S(T)$ for $R$ = Lu reveals a relatively weak temperature dependence. Even in LuPtBi, the position of the broad peak in $S(T)$ is well correlated with the broad hump in $\rho(T)$, shown more clearly in the inset of Fig. \ref{Fig5} (e) below.

The resistivity and TEP in this family depend strongly on magnetic field, exhibiting significant variations for all temperatures measured, as shown in Figs. \ref{Fig5} (a) - (f) for $R$ = Gd, Dy, and Lu. Above $T_{N}$, the observed MR behavior in this series deviates from conventional, quadratic, field dependence even up to 300 K. The maximum variation of MR for GdPtBi is clearly seen in Fig. \ref{Fig5} (a), plotted using the typical definition of MR [$\rho(H) - \rho(0) / \rho(0)$], where $\rho(0)$ is the zero field resistivity for any given temperature. A large and nearly linear MR effect of $\sim$ 800 \% can be obtained around 150 K with reduced MR for both higher and lower temperatures. TEP data as a function of field at selected temperatures are plotted in Fig. \ref{Fig5} (b). The TEP data of GdPtBi are consistent, in a qualitative way, with the behavior observed in MR. As magnetic field increases, a greatly enhanced TEP is seen around 150 K, whereas TEP depends weakly on magnetic field at 2 K and 300 K. It should be noted that the maximum value of TEP around 150 K is about $\sim$ 100 $\mu$V/K at 140 kOe. Such a large change of TEP under magnetic field is rare among non-magnetically ordered compounds and only reported in few materials such as doped-InSb and Ag$_{2+\delta}$Te \cite{Heremans2001, Sun2003}. The insets of Figs. \ref{Fig5} (a) and (b) show the $\rho(T)$ and $S(T)$ for GdPtBi at $H$ = 0, 70, and 140 kOe, respectively, as representative data sets in this series. As magnetic field increases, the position of the broad local maxima (and the sign reversal observed in zero-field) shift to higher temperature, reflecting a close correlation between resistivity and TEP.

A strong MR effect up to room temperature is also observed in DyPtBi as shown in Fig. \ref{Fig5} (c). The field dependences of the resistivity for DyPtBi are similar to GdPtBi, but no sign change is detected in Hall coefficient and TEP measurement. Note that the MR curve at $T$ = 2 K reveals anomalies corresponding to the metamagnetic transitions. As was the case for GdPtBi, the position of the broad maximum observed in zero field (around 60 K) shifts to higher temperature at $H$ = 140 kOe (inset). The TEP curves for DyPtBi, measured as a function of temperature and magnetic field, are plotted in Fig. \ref{Fig5} (d). The TEP curves as a function of field show complex field dependences for all temperatures measured. In zero field, the TEP is positive throughout the entire temperature range and has a positive maximum below 100 K (inset). As magnetic field increases the peak position moves to higher temperature. In a qualitative way, this maximum is consistent with the broad feature observed in resistivity and Hall resistivity. 

\begin{figure}
\centering
\includegraphics[width=1\linewidth]{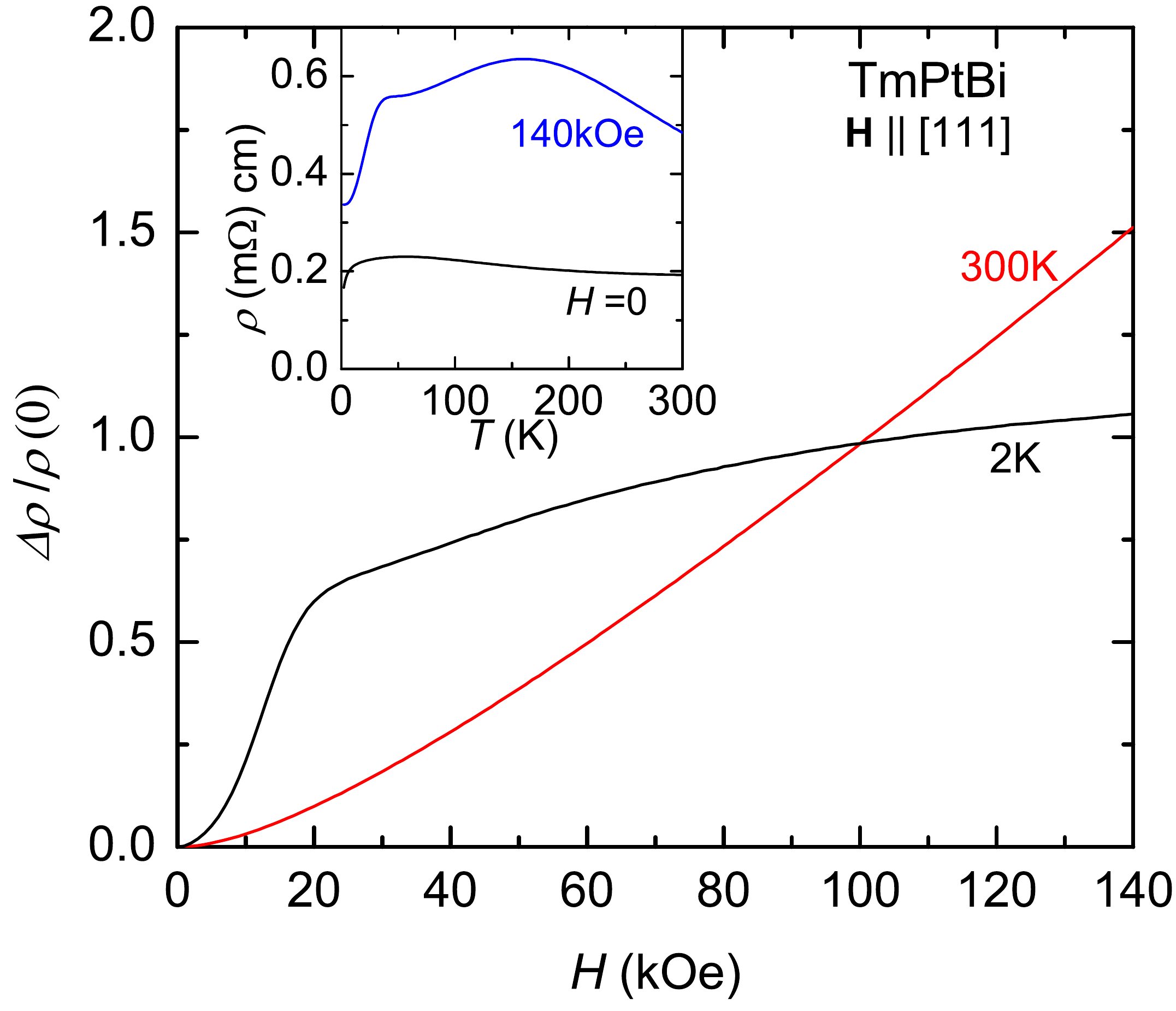}
\caption{Magnetoresistance, $\Delta \rho/\rho(0)$, of TmPtBi as a function of magnetic field at $T$ = 2 and 300 K. Inset shows electrical resistivity, $\rho(T)$, of TmPtBi as a function of temperature at $H$ = 0 and 140 kOe.}
\label{Fig6}%
\end{figure}

For LuPtBi, $\rho(T)$ curves for both $H$ = 0 and 140 kOe reveal a metallic behavior below 300 K as shown in the inset of Fig. \ref{Fig5} (e). However, $\rho(T)$ at $H$ = 140 kOe indicates a tendency of forming a broad maximum around room temperature, which consistent with TEP data. The TEP data for LuPtBi as a function of temperature and magnetic field are plotted in Fig. \ref{Fig5} (f). The TEP curve at $H$ = 140 kOe indicate a broad maximum around 250 K. In high field regime, the MR effect at 300 K is larger than that at 2 K and the MR follows a linear field dependence above $\sim$ 50 kOe as shown in the Fig. \ref{Fig5} (e). For comparison, the resistivity data of TmPtBi as a function of temperature and magnetic field are plotted in Fig. \ref{Fig6}, where again the MR at 300 K is larger than that at 2 K for $H$ $\gtrsim$ 100 kOe.

It should be noted that the MR curves for the $R$PtBi family do not collapse onto a single curve in a Kohler's plot, indicating that a single scattering process is not dominant in these Half-Heusler compounds.

\begin{figure}
\centering
\includegraphics[width=1\linewidth]{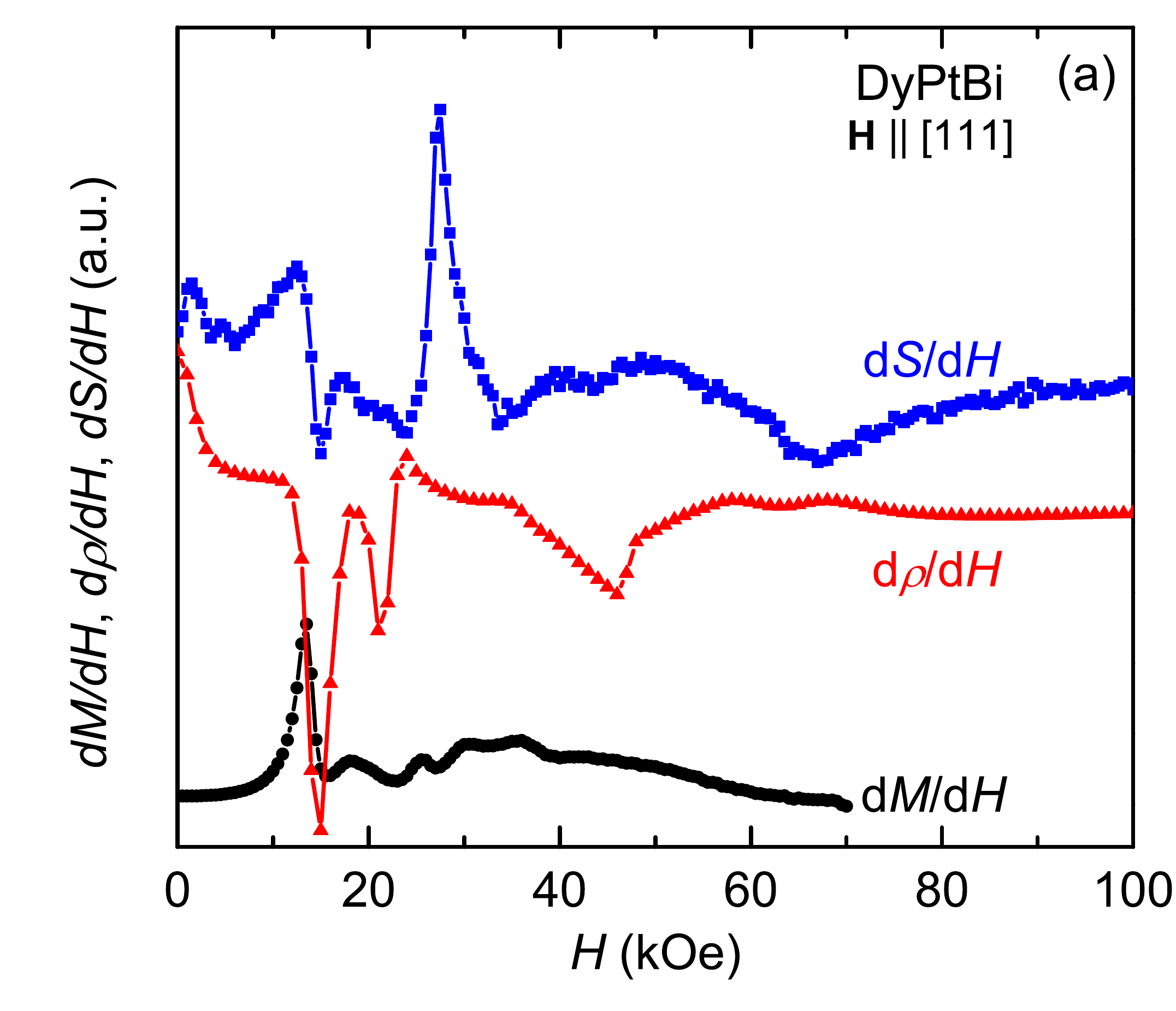}
\includegraphics[width=1\linewidth]{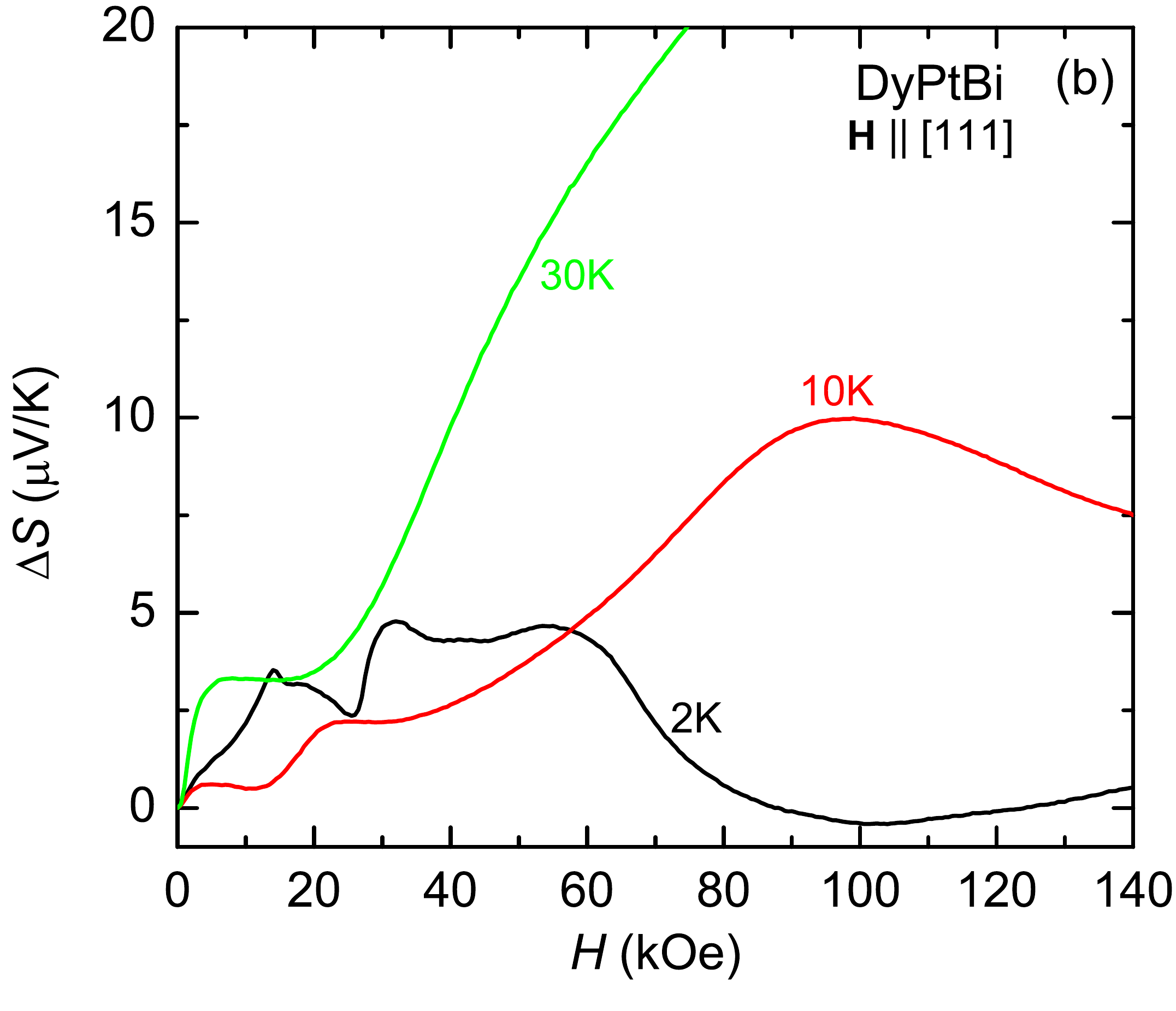}
\caption{(a) $dM/dH$, $d\rho/dH$, and $dS/dH$ curves for DyPtBi at $T$ = 2 K. (b) $\Delta S$ curves for DyPtBi at $T$ = 2, 10, and 30 K.}
\label{FigDy}%
\end{figure}

\section{Discussion}

Clear metamagnetic transitions on DyPtBi are observed in magnetization (Fig. \ref{Fig2} b), resistivity (Fig. \ref{Fig5} c), and thermoelectric power (Fig. \ref{Fig5} d) measurements at $T$ = 2 K. A sharp increase of magnetization is observed around $H_{1}$ $\sim$ 13 kOe and multiple subsequent steps are shown above $H_{1}$. These transitions are clearly indicated in the $dM/dH$ curve as shown in Fig. \ref{FigDy} (a). For comparison $d\rho/dH$ and $dS/dH$ curves at 2 K are plotted in Fig. \ref{FigDy} (a). The $H_{1}$ is clearly indicated in the derivative curves as a peak structure. Above $H_{1}$, DyPtBi has multiple metamagnetic transitions that can be detected in thermodynamic and transport measurements. Two anomalies around $\sim$46 kOe and $\sim$ 67 kOe are detected in $d\rho/dH$ and $dS/dH$ curves, whereas no signature is shown in $dM/dH$ curve. Thus, these higher field signatures may be associated with subtle changes in magnetic structure that still change the electronic structure. The TEP measurements clearly show an oscillatory behavior for both $T > T_{N}$ and $T < T_{N}$ as shown in Fig. \ref{FigDy} (b). Since the $\Delta S$ curve at 10 K shows a clear quantum oscillations, it is naturally expected that the $\Delta S$ curve at 2 K is comprised of the combination of metamagnetic transitions and the quantum oscillations. Because of the large background and small amplitude of the oscillation, the oscillatory behavior is smeared out, instead showing a step like feature below 20 kOe. Note that the oscillations are smeared out in MR measurements due to the large background. The different amplitude and frequency of the oscillations between 10 K and 2 K may reflect a different Fermi surface due to the possible formation of magnetic superzone gap (see the inset of the Fig. \ref{Fig3}). Further detailed measurements of $M(H)$, $\rho(H)$, and $S(H)$ at different temperatures and possibly higher fields are necessary to distinguish between metamagnetic transitions and quantum oscillations, which are beyond the scope of this paper.

\begin{figure}
\centering
\includegraphics[width=1\linewidth]{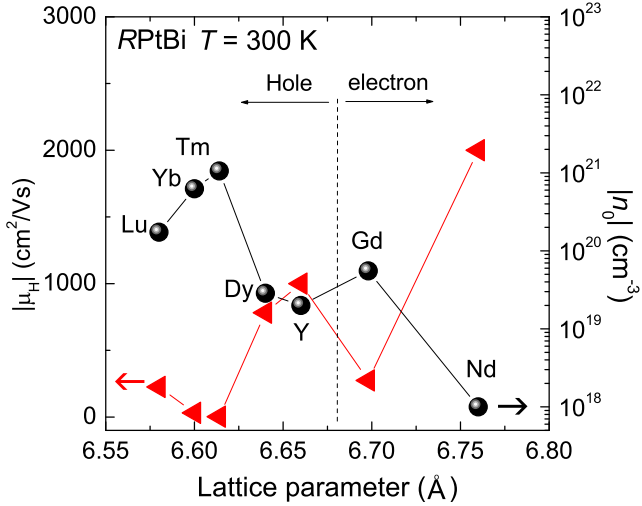}
\caption{Effective carrier density $n_{0}$ (right axis) and mobility $\mu_{H}$ (left axis) of $R$PtBi compounds at 300 K, plotted as a function of lattice constant. For $R$ = Gd and Nd, both $n_{0}$ and $\mu_{H}$ are negative. The data for $R$ = Yb, Y, and Nd are taken from the Refs. \cite{Mun2013, Butch2011, Morelli1996}.}
\label{Fig7}%
\end{figure}

Within the one-band approximation, the effective carrier density of GdPtBi is $| n_{0} |$ $\sim$ 5.5$\times$ 10$^{19}$ cm$^{-3}$ at 300 K (electron-like) and $| n_{0} |$ $\sim$ 1$\times$ 10$^{17}$ cm$^{-3}$ at 2 K (hole-like). It should be noted that since $S(T)$ and $R_{H}$ of GdPtBi indicate a sign reversal around 120 K, the $n_{0}$ estimated from the $R_{H}$ is an average of the holes and electrons weighted by their mobilities. The Hall mobility in this series is estimated by using a formula, $\mu_{H} = R_{H}/\rho$, ranging from $\sim$ 800 cm$^{2}$/Vs for $R$ = Dy at 300 K to $\sim$ 3 cm$^{2}$/Vs for Tm at 300 K. Variations of the $n_{0}$ and $\mu_{H}$ through $R$PtBi series at 300 K are plotted in Fig. \ref{Fig7} as a function of lattice parameter. In a qualitative way, the $\mu_{H}$ decreases and the $n_{0}$ increases across lanthanide contraction (lattice constant becomes smaller as the rare-earth element varies from La to Lu). Although the variation with lattice parameter shown in Fig. \ref{Fig7} is only a rough trend, the $R$ = Lu and Y members appear near Yb/Tm and Dy/Gd members. This rules out any significant role of the rare-earth local moment and rather indicates that the gross trend across the series is steric rather than magnetic in origin. If magnetism were the key variable, then a global maximum near Gd-Ho would appear rather than a general rise with decreasing unit cell volume.

The transport properties in this series are governed by hole-like carrier for $a$ $\lesssim$ 6.68 \AA ~and by electron-like carrier for $a$ $\gtrsim$ 6.68 \AA, which indicates an altering of carrier pocket sizes by moving chemical potential. In particular, for $R$ = Gd, $|\mu_{H}|$ is $\sim$ 300 cm$^{2}$/Vs at 300 K, whereas the extremely high mobility of $\sim$ 20,000 cm$^{2}$/Vs at 2 K is obtained. This high value of mobility can be produced by a zero band gap and linear dispersion of bands with low effective mass ($m^{*}$), which is found to be comparable to that of single crystals topological insulators \cite{Thio1998, Butch2010}.

The observed transport properties of GdPtBi are very similar to those of the silver chalcogenides, where the maximum MR effect is detected at the sign change of $R_{H}$ due to the band crossing \cite{Xu1997, Abrikosov1998}. Such sign reversals in $R_{H}$ and TEP are expected when the electron and hole bands cross and the maximum MR effect is expected at the crossing point because the $n_{0}$ approaches zero and the $m^{*}$ becomes smaller ($\rho \propto m^{*}/n_{0}$). If the large MR in this family is due to the balancing two-types of carriers, then lattice parameter values around 6.68 \AA ~should give a maximum MR (Fig. \ref{Fig7}). We note that the sign change of $R_{H}$ is considered to be a necessary condition to produce a large, linear MR effect in silver chalcogenides, whereas $R$PtBi compounds reveal that a large and nearly linear MR can be produced for both cases with sign reversal ($R$ = Gd) and without sign change ($R$ = Dy) in $R_{H}$ and TEP.

One possible way of rationalizing the behavior of the transport properties of the $R$PtBi series is to focus on GdPtBi as the archetype. Based on Ref. \cite{Abrikosov1998}, when the chemical potential, $\mu$, is lying at the crossing point of liner bands in a gapless semiconductor, the Hall coefficient and TEP are zero due to the electron-hole symmetry and a large linear MR can be produced. At the crossing point, the Fermi surface is a point giving rise to a linear dispersion. At high temperatures, thermally excited electrons produce electron-type carriers in GdPtBi. As temperature decreases and $\mu$ passes through the crossing point, there is a sign change of $R_{H}$ and TEP around 120 K. When a magnetic field is applied, the $\mu$ also lowers and moves away from the crossing point into the hole-type regime. In this regard, negative $R_{H}$ in low field becomes positive in high fields and the sign reversal in $S(T)$ moves toward higher temperature as magnetic field increases. In a similar manner, when the lanthanide contraction is considered, the $\mu$ located above the crossing point for lighter rare-earth moves lower than the crossing point for heavier rare-earth, continuously moving toward more hole-like system (Fig. \ref{Fig7}). Clearly, pressure tuning of $R$PtBi compounds will provide further opportunities to manipulate the band structure and optimize the response of transport properties in this picture. To some extent such tuning can also be accomplished with mixed rare earth occupancy on the $R$-site \cite{Canfield1991}.

\begin{figure}
\centering
\includegraphics[width=1\linewidth]{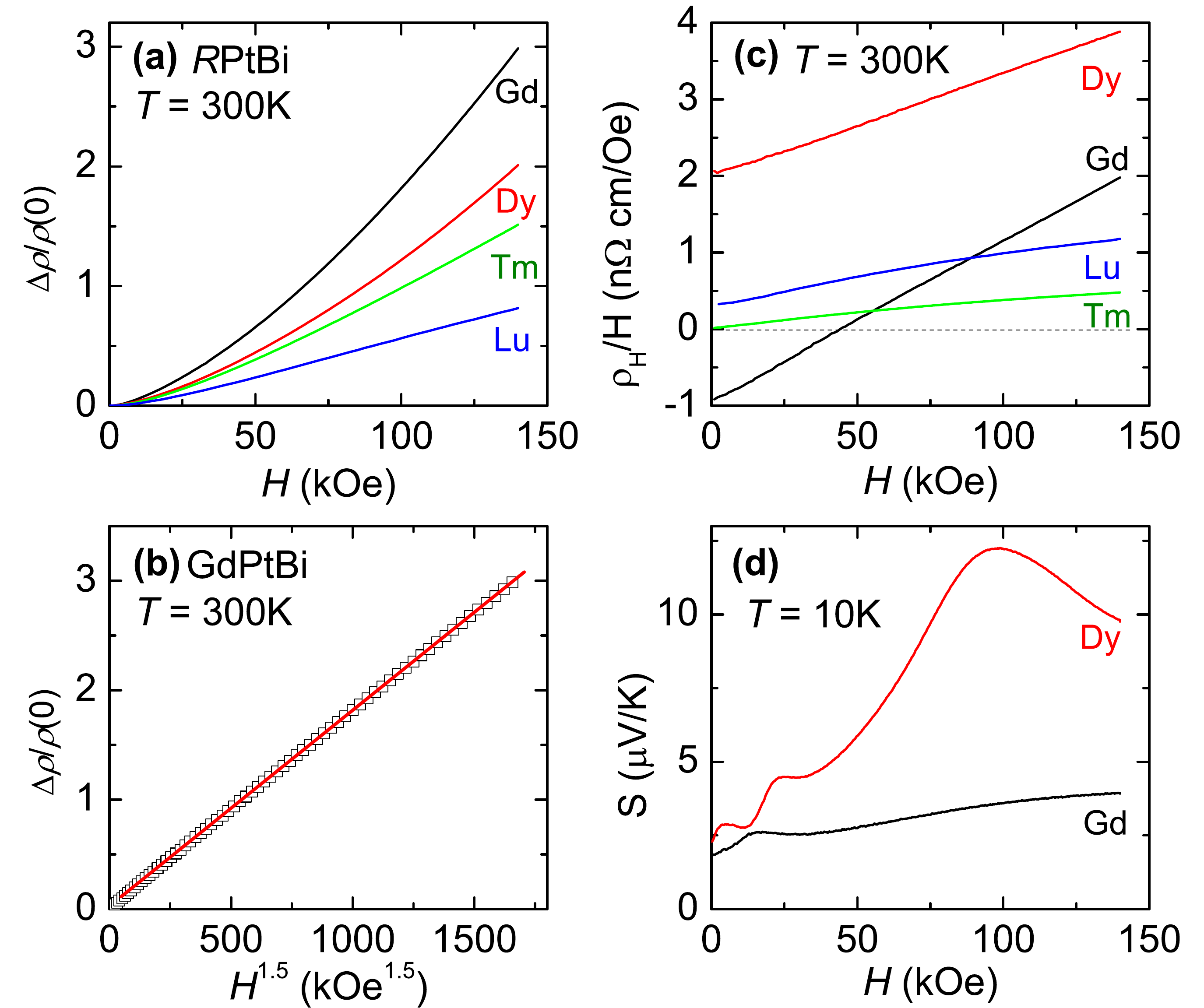}
\caption{(a) Magnetoresistance of $R$PtBi ($R$ = Gd, Dy, Tm, and Lu) at 300 K. (b) MR of GdPtBi, plotted as a function of $H^{1.5}$. The solid-line is the guide to the eyes. (c) $R_{H}$ of $R$PtBi at 300 K. (d) $S(H)$ for GdPtBi and DyPtBi at 10 K.}
\label{Fig8}%
\end{figure}

Figure \ref{Fig8} (a) shows the large, room temperature MR effect observed in the $R$PtBi compounds. The large MR effect decreases as the rare-earth is traversed from Gd to Lu. The MR curves show an approximately $H^{2}$ dependence only in the very low field regime, whereas in the high field regime, the MR curves vary from an approximately $H^{1.5}$ dependence for $R$ = Gd (Fig. \ref{Fig8} (b)) to a nearly linear field dependence for $R$ = Lu (see Fig. \ref{Fig5} (e)). For the non moment bearing LuPtBi compound, the quadratic MR can be well understood from conventional mechanism, whereas the origin of the linear MR is currently unknown.

In general, because the mobility in ordinary metals decreases as temperature rises, the $H^{2}$-dependence of MR effect is larger at low temperatures and negligible (changing only few percent) at high temperatures. The size of the MR in the $R$PtBi compounds is comparable to or smaller than ordinary metals at low temperatures, but much larger than those at high temperatures. If it is caused by mobility fluctuations, as suggested for the cases of doped-InSb and Cd$_{3}$As$_{2}$ \cite{Hu2008, Narayanan2015}, then the MR effect should continuously decrease as temperature increases. The MR effect observed in $R$PtBi compounds depends weakly on temperature, varying by roughly a factor of $\sim$ 5 between 2 and 300 K. Therefore, the anomalous MR effect in this series (deviating from conventional $H^{2}$ dependence) cannot be explained by the classical model governed by mobility. LuPtBi is particularly of interest in the high field region, where the MR effect at 300 K is larger than that at 2 K, implying almost temperature independent MR effects.

Driving the GdPtBi system away from the band crossing region via temperature, magnetic field, and lanthanide contraction reduces the magnitude and changes the functional form of the MR. At 300 K, $\rho_{H}/H$ curves for $R$ =  Gd, Dy, Tm, and Lu are plotted in Fig. \ref{Fig8} (c), where $\rho_{H}/H$ can be approximated as being linear in field. It should be noted that the $\rho_{H}$ of $R$ = Yb in this family is linear in field above 100 K \cite{Mun2013}, whereas the $\rho_{H}$ of other rare-earth compounds varies as $\sim H^{2}$. This strong deviation from linear field dependence of $\rho_{H}$, for whole temperature regime measured, excludes the generally applicable analysis \cite{Pippard1989}.

Although there is currently no direct evidence of linearly dispersive band structures in the $R$PtBi compounds, such features should not be excluded as possible contributors to the anomalous MR. Recently exceptionally large MR has been associated with such features in WTe$_{2}$, Cd$_{3}$As$_{2}$, and PtSn$_{4}$ \cite{Mun2012, Narayanan2015, Ali2014}. Previous ARPES experiments on $R$PtBi compounds clearly observed the conducting surface electronic structure, where several surface bands with spin-orbit splitting cross the chemical potential (Fermi level) \cite{Liu2011}. Among those bands the $\Gamma$ pockets are particularly interesting. The size of $\Gamma$ pockets in $R$PtBi becomes bigger as the rare-earth is traversed from Gd to Lu, implying a systematic variation of effective carrier occupancy through members of this half-Heusler family.

The tiny size of $\Gamma$ pockets can also be inferred from quantum oscillations. For GdPtBi and DyPtBi, a clear oscillatory behaviors at 10 K (above $T_{N}$) are observed in TEP measurements as shown in Fig. \ref{Fig8} (d). The oscillation frequency of DyPtBi is estimated to be $\sim$ 0.4 T by considering low field maximum and minimum. The third maximum is not periodic in a 1/$H$ plot, probably approaching quantum limit before third peak due to the spin splitting effects. Although the frequency of GdPtBi cannot be calculated exactly, it can be approximated to be much smaller than that of DyPtBi by considering the first maximum, which is consistent with the evolution of $\Gamma$ pockets. Previously, similar oscillations in NdPtBi with 0.74 T are observed in MR below 30 K \cite{Morelli1996}. The systematic change of transport properties in this series may also be explained qualitatively by considering both bulk bands and the surface bands with linear dispersion. In this respect, it is necessary to identify both bulk and actual surface band contributions further by high resolution ARPES and angular dependence of quantum oscillations to elucidate the non-trivial MR effect in this series. It has been shown that the dimensionality plays a crucial role in materials with non-trivial topology. When the thickness is reduced, the topologically protected surface state become dominant in transport, further transport measurements on thin film LuPtBi will also illuminate the origin of linear contribution of MR.

\section{Summary}

In summary, we have studied the transport properties of the $R$PtBi Half-Heusler compounds. It is found that: i) the MR depends weakly on temperature, which is quite different from what has been observed in ordinary metals; ii) a conventional scattering process, where the Kohler's rule holds, is not dominant; iii) the MR and mobility vary with temperature in different manners; iv) the high field MR, at 300 K, does not saturate and indicates a deviation from the quadratic field dependence; vi) lastly, the unusually large MR effect at room temperature can be tuned systematically through lanthanide contraction. This study provides a solid platform to control MR and a way to engineer the size of MR at room temperature without introducing disorder, unlike other materials such as silver chalcogenides, doped-InSb, and topologically non-trivial materials.

\begin{acknowledgments}
This work was supported by the U.S. Department of Energy, Office of Basic Energy Science, Division of Materials Sciences and Engineering. The research was performed at the Ames Laboratory. Ames Laboratory is operated for the U.S. Department of Energy by Iowa State University under Contract No. DE-AC02-07CH11358. The work at Simon Fraser University was supported by Natural Sciences and Engineering Research Council of Canada. 
\end{acknowledgments}

\end{document}